\documentclass[prl,twocolumn]{revtex4}
\usepackage{graphicx}
\usepackage{amsmath,mathtools,amssymb,braket,float,bm,color,physics}
\usepackage{comment}

\begin{document}

\title{
Convection of mono-disperse particles in a highly filled rotating cylinder}
\author{Shoichi Yoneta}
\author{Hiroyuki Ebata}
\author{Shio Inagaki}
\affiliation{Dept. of Physics, Kyushu University}


\begin{abstract}
	We investigate the occurrence of spontaneous convection in a coaxial cylinder highly filled with mono-disperse spheres.
	To analyze the flow field non-invasively, initial pulses consisting of colored particles are placed at equal intervals.
	By analyzing the spatio-temporal distribution of these pulses, we obtained axial velocity profiles for both the surface and subsurface regions.
	Our advection-diffusion equations with steady advection terms incorporate experimentally obtained axial velocity profiles in the surface layer, while the rest of the components are estimated using azimuthal symmetry and volume conservation.
	The validity of our model is confirmed by comparing experimental data with numerical solutions for both the spatio-temporal distribution and cross-sectional profile of the colored particles.
\end{abstract}

\maketitle

Granular materials exhibit a rich range of behaviors, including melting, flowing, and even boiling, in response to applied mechanical agitation \cite{Jaeger96:review}.
Unlike traditional states of matter such as solid, liquid, or gas, granular materials defy simple classification due to the dissipative interactions between their constituent particles.
Granular flow has been extensively studied as it is widely found both in nature and industrial processes
\cite{GdR:review,Pouliquen08:review,Iverson97:DebrisFlow}.

Granular materials exhibit a fascinating property of segregation when subjected to mechanical agitation, such as rotation and vibration \cite{Thomas2011:Review}.
When a cylinder is highly filled with a bi-disperse granular mixture and rotated horizontally, segregated pattern emerges as the particles flow down the surface \cite{Inagaki10:RotatingDrum}.
One may expect that the bulk under the fluidized region would be frozen where geometrical restrictions prevent the independent motion of the particles. However, granular materials can flow collectively  over extended time scales \cite{Komatsu2001:CreepMotion}.
Reflecting this highly viscous fluid-like nature, very slow three-dimensional convective motion has also been observed in such highly filled bidisperse systems 
\cite{Inagaki10:RotatingDrum,Nakagawa1998:RotatingDrum,Kuo2006:RotatingDrum}.
In the case of a coaxial cylinder, 
the segregated pattern moves either in one direction or oscillatively depending on the slight change in the fill level  \cite{Inagaki15:bidisperse}.
Rietz and Stannarius also observed
convection rolls in horizontally rotating cells \cite{Stannarius08:RotatingCell}.
One question might arise here as to whether the motion of the segregated pattern is originated from the rearrangements caused by the segregation or a confined system filled with discrete particles can spontaneously form collective motion, such as convection.

To understand three-dimensional granular flow, there are several studies to visualize internal structures in both invasive and non-invasive manners.
The simplest invasive method is to open the container with or without solidifying particles and observe the cross-section \cite{Santomaso04:solid,Kuo2006:RotatingDrum,Inagaki10:RotatingDrum}. Although this method is simple and costless in labor and time, we cannot observe a time evolution of subsurface convective motion. In this respect, non-invasive techniques such as Magnetic Resonance Imaging \cite{Hill97:MRI}, Nuclear Magnetic Resonance \cite{Nakagawa1998:RotatingDrum}, and Positron Emission Particle Tracking \cite{Ding01:PEPT} are effective, although they will cost enormously for sufficient resolution in time and space.

\begin{figure}[b]
	\centering
	\includegraphics[width=\columnwidth]{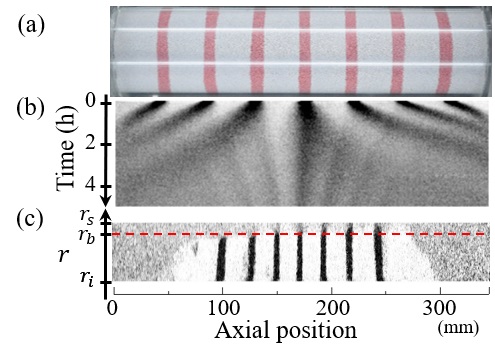}
	\caption{
	(a) Initial configuration with seven colored pulses with a fill level of 95 $\%$.
	(b) Spatio-temporal diagram of the surface pattern.
	(c) Cross-sectional image after 5 hours of rotation. The aspect ratio of the image is increased by 2.8 vertically for ease of viewing.
	}
	\label{fig:Expt.1}
	\end{figure}

The aim of the present study is to elucidate the underlying mechanisms of granular convection in a densely packed system, specifically focusing on the occurrence of convection in the absence of size segregation. 
The experiments were conducted using mono-disperse spheres with an acrylic coaxial cylinder. 
The cylinder has inner and outer radii, $r_i$ and $r_s$, respectively, measuring 20 and 40 $\rm{mm}$, with a length $L$ of 345 $\rm{mm}$. To enable direct observation of the cross-section, the cylinder was divided into two semicircular sections and securely fastened together, allowing for separation after rotation.
We used spherical alumina beads (280±70 $\rm{\mu m}$, density $\rm{g/cm^3}$). To visualize the flow field, we used white and red particles. The red particles were colored with oil-based magic ink.
We placed the cylinder vertically, and poured alternately white and colored particles into the cylinder to achieve a fill level of 95 $\%$.
The fill level was defined as the particle volume normalized by $64\%$ of the cylinder volume (Random Close Packing \cite{Jaeger92:RCP}). The cylinder was placed on parallel shafts and rotated at 15 $\rm{rpm}$ using a brushless DC electric motor (Oriental Motors). 

To capture the overall picture of the flow,
we carried out an experiment with seven colored pulses placed at equal intervals in the initial state (Fig. \ref{fig:Expt.1}(a)). 
The cylinder was rotated for 5 hours.
The images are grayscaled for analysis.
Figure \ref{fig:Expt.1}(b) shows the spatio-temporal diagram of the surface flow field, created by stacking one-pixel-high horizontal lines of the images taken every 30 seconds.
The diagram exhibits distinct trajectories of the seven pulses, originating from the initial position and extending towards both end-walls. This observation strongly suggests the occurence of flow from the center to the end-walls on the surface of the cylinder.
Remarkably, additional blurred trajectories were observed originating from the initial position of the pulses and extending towards the center of the cylinder, counter to the predominant motion of the prominent pulses.

Figure \ref{fig:Expt.1}(c) is one of the two cross-sections obtained after rotating the coaxial cylinder for 5 hours.
The upper and lower sides of the rectangular cross-section corresponds to the outer and inner cylinder.
In the central region, the seven pulses of colored particles remained vertically aligned, indicating that the axial velocity in this region is independent of the radial direction. At the boundary of the inner cylinder, slip occurs, suggesting plug flow characteristics similar to Bingham fluids that exhibit yield stress.
The intervals between the pulses remain nearly equal but narrower than the initial configuration.
Furthermore, there is a one-to-one correspondence between the position of the colored pulses of the cross-section and the final position of the blurred trajectories on the spatio-temporal diagram.
It indicates that the blurred trajectories are colored particles that exude from the subsurface. The time evolution of the positions of the colored pulses in the subsurface region was detected in a non-invasive manner by tracking the blurred trajectories.

Based on the cross-sectional image in Fig. \ref{fig:Expt.1}(c), we delineate two distinct layers within the system: a surface layer characterized by particles traversing the fluidized region under rotation, and an inner layer where particles exhibit collective bulk movement.
Near the center of the cylinder, there is a clear horizontal boundary between the mixed and unmixed domains. The mixed domain at the surface is resulted from mixing of the particles while they flow down the fluidized region. By assuming the thickness of the surface layer is constant irrespective of the axial position, we determine the layer boundary as depicted by a dashed line in Fig. \ref{fig:Expt.1}(c). In the following, the surface and inner layers refer to the regions above and below the dashed lines, respectively.

\begin{figure}[tb]
	\centering
	\includegraphics[width=\columnwidth]{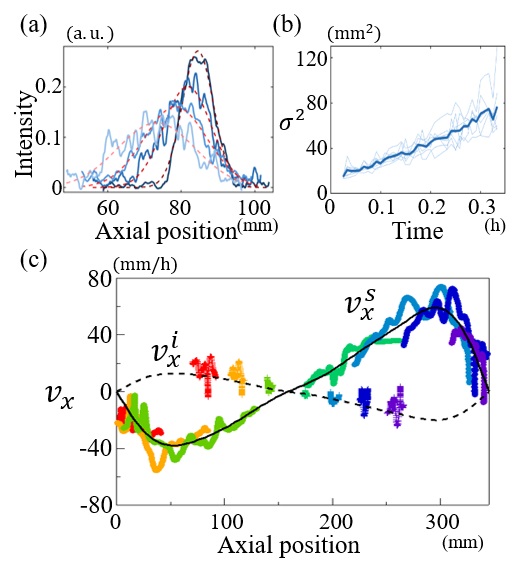}
  \caption{
	(a) Intensity of the second pulse from the left in the spatio-temporal diagram at time $t=3, 11, 19, 26 \, \rm{min}$ from dark blue to lilght blue. Dashed lines denote the Gaussian fitting curves.
	(b) Variance of the Gaussian fitting of the pulse intensity (light blue: seven individual pulses, dark blue: average of seven pulses). 
	(c) Axial velocity profiles in the surface (o) and inner (*) layers. 
  Solid line: the fitting curve of $v_x^s$.
	Dashed line: the estimated axial velocity in the inner layer.
  The pulses from left to right corresponds to the data with colors from red to purple.
	}
	\label{fig:pulse}
\end{figure}
Figure \ref{fig:pulse}(a) demonstrates the time evolution of the intensity of the second pulse from the left in the spatio-temporal diagram in Fig. \ref{fig:Expt.1}(b). The pulse center moves leftward and its width increases progressively. 
This behavior is consistent among the other six pulses, indicating symmetry at the cylinder's center. The observations suggest an advection flow from the center to the end walls, accompanied by particle diffusion. The intensity of each pulse is fitted with a Gaussian distribution, and its variance, representing the pulse width, is plotted against time in Fig. \ref{fig:pulse}(b). The thick line represents the average variance. By assuming 1-dimesional normal diffusion, the axial diffusion coefficient is estimated to be approximately $82 \, \rm \rm{mm^2/h}$.
The axial velocities in the inner and surface layers were determined by analyzing the time evolution of the pulse centers in the spatio-temporal diagram, where the pulse intensities were fitted with a Gaussian distribution (Fig. \ref{fig:pulse}(c)). The overlapping velocities among the pulses indicate position-dependent velocity distribution. 
It should be noted that in a half-filled monodisperse system, the normal diffusion in axial direction was also observed numerically \cite{Taberlet2006:Diffusion}.

We propose an advection-diffusion model considering both advective motion and diffusive behavior in our system. Using cylindrical coordinates $(r, \theta, x)$, where $x$ is the axial position from the left end of the cylinder, $r$ is the radial distance from the $x$-axis, and $\theta$ is the azimuthal angle, we assume symmetry with respect to $\theta$ despite the presence of a narrow avalanche along the axis. Additionally, we consider that the colored particles follow the velocity field without influencing it, making the total particle system's velocity equivalent to that of the colored particle component.
By combining the assumption of incompressibility and axial symmetry, along with the law of conservation of mass,
we can derive the advection diffusion equation 
in terms of the concentration of the colored particles, $\phi_A$, which is defined as the ratio of the density of the colored particles to the density of the total particle system as follows:
\begin{eqnarray} 
	\pdv{\phi_A}{t}+v_x^\alpha\pdv{\phi_A}{x}+v_r^\alpha\pdv{\phi_A}{r} = 
	\dfrac{1}{r}\pdv{}{r} \qty[D_{r}^\alpha r\pdv{\phi_A}{r}] + D_{x}^\alpha \pdv[2]{\phi_A}{x}		
	\label{eq:2Dade}
\end{eqnarray}
where the diffusion coefficients in axial and radial directions are presented by $D_x^\alpha$ and $D_r^\alpha$, while
the axial and radial velocities within the total particle system are denoted as $v_x^\alpha$ and $v_r^\alpha$, respectively.
The superscript letter $\alpha$ represents $i$ for the inner layer and $s$ for the surface layer. 
We set the diffusion coefficients in the inner layer, $D_x^i$ and $D_r^i$, to zero
because the cross-sectional image in Fig. \ref{fig:Expt.1}(c) clearly demonstrates the absence of diffusion 
within the inner layer. 
The comprehensive derivation is given in Supplemental Material A.

\begin{figure}[tb]
	\centering
	\includegraphics[width=\columnwidth]{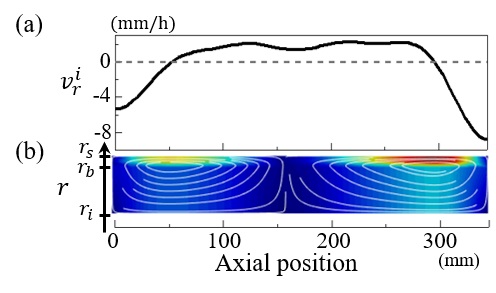}
  \caption{
	(a) Estimated radial velocity profile $v_r^i(x,r=r_b)$ derived from Eq. (\ref{eq:vxi}).
	(b) Stream line with heatmap.
	The aspect ratio is increased by 2.8 vertically  for ease of viewing.
	}
	\label{fig:vr}
\end{figure}
To solve these equations numerically, the diffusion coefficients and the steady advection terms are required. 
First, the axial velocity in the surface layer, $v_x^s$, was fitted to the experimental data using the following function:
\begin{equation}
	v_x^s = \sum_{k=1}^4 \Big[ A_k \sin 2 k \pi \tilde{x} + B_k \cos (2 k-1) \pi \tilde{x} \Big], \label{eq:vxs}
\end{equation}
where $\tilde{x}=x/L - 1/2$.
The detalied values of $A_k$ and $B_k$ are listed in Supplemental Material B.
The fitting function is plotted by a solid line in Fig. \ref{fig:pulse} (c).
We assume that the axial velocities in both the surface and inner layers are independent of $r$. 
In the surface layer, the avalanche layer is sufficiently thin to assume the axial velocity is radially uniform. 
In the inner layer, we assume that the axial velocity is independent of $r$, as it was observed to be consistent 
in Fig. \ref{fig:Expt.1}(c).
Second, the axial velocity in the inner layer is derived from that in the surface layer. 
Incompressibility implies constant density and no inflow or outflow in the system. Therefore, the net flux through any cross-section must be zero.
Given these conditions, $v_x^i$ and $v_x^s$ must satisfy the relation 
$\pi (r_b^2-r_i^2) v_x^i+\pi (r_s^2-r_b^2) v_x^s = 0$.
The radial position of the layer boundary, $r_b$, is set at $3.6 \, \rm{mm} $, determined from the cross-sectional image in Fig. \ref{fig:Expt.1}(c).
Consequently, the axial velocity in the inner layer, $v_x^i$, can be  derived from the axial velocity in the surface layer, $v_x^s$, as
\begin{equation}
v_x^i=-\frac{r_s^2-r_b^2}{r_b^2-r_i^2} \, v_x^s. \label{eq:vxi}
\end{equation}
The estimated $v_x^i$ is represented by the dashed line in Fig. \ref{fig:pulse}(c), which shows a good agreement with the experimental data.

In terms of the radial velocity profiles, 
based on the assumption of incompressibility and axial symmetry and the law of conservation of mass, the inflow and outflow of particles in the axial direction must be balanced with that in the radial direction.
As a result, the radial velocity distributions in the inner and surface layers, $v_r^i$  and $v_r^s$, can be written as
\begin{eqnarray}
	v_r^\alpha(x,r) =\frac{{r_\alpha }^2-r^2}{2r}\dv{v_x^\alpha (x)}{x}.  \qquad (\alpha = i, s)  \label{eq:vr}
\end{eqnarray}
For a detailed derivation, refer to Supplemental Material C.
By substituting Eqs. (\ref{eq:vxs}) and (\ref{eq:vxi}) into Eq. (\ref{eq:vr}), we can estimate the radial velocity profile in the surface and inner layers.
It should be noted that the radial velocity $v_r^\alpha$ depends on both $r$ and $x$.
The radial velocity profile $v_r^i(x, r=r_b)$ is depicted in Fig. \ref{fig:vr}(a).
By utilizing the experimentally obtained $v_x^s$, we derive the analytical expressions for the remaining components, $v_x^i$, $v_r^s$, and $v_r^i$, by employing the principles of volume conservation and axial symmetry.
Based on the overall velocity profiles shown in Figs. \ref{fig:pulse}(c) and \ref{fig:vr}(b), the stream line is depicted with a heatmap.
We inferred the presence of a two-vortex internal structure.
Specifically, the right half exhibits a clockwise vortex, while the left half exhibits a counterclockwise vortex.

\begin{figure}[t]
 \begin{center}
	\includegraphics[width=\columnwidth]{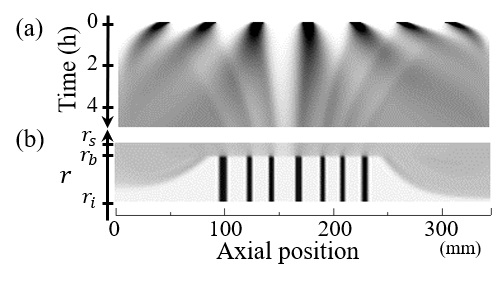	}
  \caption{Numerical solution of the advection-diffusion equation.
	The color corresponds to the concentration of the colored particles.
	(a) Spatio-temporal diagram with a duration of 5 hours.
	(b) Cross section.
	The aspect ratio of the image is increased by 2.8 vertically for ease of viewing.}
  \label{fig:simu}
 \end{center}
\end{figure}

To validate the advection-diffusion equations in Eq. (\ref{eq:2Dade}), 
we numerically solved them by using an explicit finite difference method.
The diffusion coefficients in axial and radial directions in the surface layer, $D_x^s$ and $D_r^s$, were set at $30 \, \rm{mm^2/h}$.
To ensure the agreement between the numerical solution and the experimental results, the diffusion coefficient in the surface layer, $D_x^s$, was adjusted to a lower value than initially estimated from the pulse data in Fig. \ref{fig:pulse}(b). 
The detail of the numerical simulation is given in Supplemental Material D.
The numerical solution of the advection-diffusion equations described in Eq. (\ref{eq:2Dade}) is presented in Fig. \ref{fig:simu}.
The spatio-temporal plot successfully reproduces the qualitative characteristics observed in the experiment, such as the clear trajectories in the surface layer and the diffuse trajectories exuding from the inner layer shown in Fig. \ref{fig:Expt.1}(c).
However, the pulses in the spatio-temporal diagram of the numerical solution appear to move slightly slower than those observed in the experiment, indicating a potential underestimation of the velocity field in the model.
The cross-sectional features in the experiment shown in Fig. \ref{fig:Expt.1}(c) are also well represented in the numerical solution shown in Fig. \ref{fig:simu}(b).

When a container is nearly completely filled with a bi-disperse granular mixture and rotated horizontally, three dimensional convective motions were observed
\cite{Nakagawa1998:RotatingDrum,Kuo2006:RotatingDrum,Inagaki10:RotatingDrum,Inagaki15:bidisperse}.
Our current study shows that in a densely packed granular system, convective motion can occur even without segregation.
It indicates that the segregated pattern was formed by the surface flow, and the collective band movement itself was attributed to the macroscopic convection.

In this paper, we have presented an experimental study of granular convection nearly filled in a coaxial cylinder.
The densely packed mono-disperse system exhibited convective motions in the absence of segregation.
We obtained the axial velocity distribution in the inner layer from the trajectories of the pulses in the spatio-temporal diagram.
We successfully derived the axial velocity in the inner layer and the radial velocity distributions in both the inner and outer layers
from the axial velocity in the surface layer
by employing the conservation law and axial symmetry.
Using these velocity profiles as steady advection terms, we numerically solved the advection-diffusion equation and successfully reproduced the experimental results.
This provides compelling evidence that the trajectories of the blurred pulses capture the flow in the inner layer in a non-invasive manner.
 
Our model does not explicitly include parameters characterising the granular medium. We assume that the physical properties of the media have been included in the advection and diffusion of the flow field and the thickness of the surface avalanche, which are observable quantities. For example, the ratio of particle diameter to the thickness of the co-axial cylinder significantly influences the fluidity of the media and, consequently, the overall flow field.
Moreover, the thickness of the surface layer is influenced by various factors, including the particle shape, the fill level and rotational speed of the cylinder.
However, a deeper understanding of how the velocity profile is determined based on the material properties of the particles requires further experimental studies.


A fundamental question still remains as to how such a stationary flow field could originally arise by rotation.
The sudden drop of $v_r^i$ near the end walls in Fig. \ref{fig:vr} (a) suggests that friction from the end walls could be a driving force behind the convective motion.
Previous experimental \cite{Santomaso04:solid,Huang2013:Endwall,Pohlman2006:Endwall,Maneval2005:Endwall} and numerical \cite{Chen2008:Endwall,Arntz2013:Endwall,DOrtona2022:Circulation} investigations have explored the influence of end walls on the flow dynamics.
Pohlman experimentally demonstrated curved streamlines adjacent to the end walls in a half-filled cylinder, where particles flow away from the end wall  in the upstream and then flow back toward the end wall in the downstream to conserve mass \cite{Pohlman2006:Endwall}. 
Since the magnitude of the axial velocity is larger in the downstream than in the upstream, each time the container rotates once and the particles experience a surface avalanche, on average, an axially outward flow is generated at the wall. Subsequently, it must turn radially inwards at the end walls to conserve mass.
This asymmetric axial flows between the upstream and downstream is one of the strongest candidates for the origin of the convection.
As a preliminary investigation, we examined the effect of rotational speed, and the results indicated that higher rotational speeds resulted in faster convection motion (see Supplemental Material E).
This observation is consistent with the experimental findings, where the magnitude of axial velocity at the end walls substantially increased with increasing rotation speed \cite{Pohlman2006:Endwall}.
Further research will undoubtedly deepen our comprehension of the underlying mechanisms governing granular convection. 

This work was supported by JSPS KAKENHI Grant Numbers 22K03468.


\begin{thebibliography}{}
\bibitem{Jaeger96:review}
H. M. Jaeger, S. R. Nagel, and R. P. Behringer, {\it{Rev. Mod. Phys.}} {\bf 68}, 1259 (1996).

\bibitem{GdR:review}
GDR MiDi, {\it{Eur. Phys. J. E}} {\bf 14}, 341 (2004).

\bibitem{Pouliquen08:review}
Y. Forterre and O. Pouliquen, {\it {Annu. Rev. Fluid Mech.}} {\bf 40}, 1 (2008).

\bibitem{Iverson97:DebrisFlow}
R. M. Iverson, {\it{Rev. Geophys.}} {\bf 35}, 245 (1997).

\bibitem{Thomas2011:Review}
G. Seiden and P. J. Thomas,
{\it{Review of Modern Physics}} {\bf 83}, 1323 (2011).

\bibitem{Inagaki10:RotatingDrum}
S. Inagaki and K. Yoshikawa, {\it{Phys. Rev. Lett.}} {\bf 105}, 118001 (2010).

\bibitem{Komatsu2001:CreepMotion}
T. S. Komatsu, S. Inagaki, N. Nakagawa, and S. Nasuno,
{\it{Phys. Rev. Lett.}} {\bf 86}, 1757 (2001).

\bibitem{Nakagawa1998:RotatingDrum}
M. Nakagawa, S. A. Altobelli, A. Caprihan, E. Fukushima
{\it{Chemical Engineering Science}} {\bf 52}, 4423 (1997).

\bibitem{Kuo2006:RotatingDrum}
H. P. Kuo, P. Y. Shih, and R. C. Hsu, {\it{AIChE J.}} {\bf 52}, 2422 (2006).

\bibitem{Inagaki15:bidisperse}
S. Inagaki, H. Ebata and K.Yoshikawa, {\it{Phys. Rev. E}} {\bf 91}, 010201(R) (2015).

\bibitem{Stannarius08:RotatingCell}
F. Rietz and R. Stannarius, {\it{Phys. Rev. Lett.}} {\bf 100}, 078002 (2008);
{\it{Phys. Rev. Lett.}} {\bf 108}, 118001 (2012).

\bibitem{Santomaso04:solid}
A. Santomaso, M. Olivi, and P. Canu,
{\it{Chem. Eng. Sci.}} {\bf 59}, 3269 (2004).

\bibitem{Hill97:MRI}
K. M. Hill, A. Caprihan, and J. Kakalios, {\it{Phys. Rev. Lett.}} {\bf 78}, 50 (1997).

\bibitem{Ding01:PEPT}
Y. L. Ding, J. P. K. Seville, R. Forster and D. J. Parker, {\it{Chem. Eng. Sci.}} {\bf 56}, 1769 (2001).

\bibitem{Jaeger92:RCP}
H. M. Jaeger and S. R. Nagel,
{\it{Science}} {\bf 255}, 1523 (1992).


\bibitem{Taberlet2006:Diffusion}
N. Taberlet and P. Richard,
{\it{Phys. Rev. E}} {\bf 73}, 041301 (2006).



\bibitem{Pohlman2006:Endwall}
N. A. Pohlman, J. M. Ottino, and R. M. Richard,
{\it{Phys. Rev. E}} {\bf 74}, 031305 (2006).

\bibitem{Huang2013:Endwall}
A.-N. Huang, L.-C. Liu, and H.-P. Kuo
{\it{Powder Technology}} {\bf 239}, 98 (2013).

\bibitem{Maneval2005:Endwall}
J. E. Maneval, K. M. Hill, B. E. Smith, A. Caprihan, E. Fukushima,
{\it{Granular Matter}} {\bf 7}, 199 (2005).

\bibitem{Chen2008:Endwall}
P. Chen, J. M. Ottino, and R. M. Lueptow,
{\it{Phys. Rev. E}} {\bf 78}, 021303 (2008).

\bibitem{DOrtona2022:Circulation}
U. D'Ortona, N. Thomas, and R. M. Lueptow,
{\it{Phys. Rev. E}} {\bf 97}, 052904 (2018); {\bf 105}, 014901 (2022).


\bibitem{Arntz2013:Endwall}
M. M. H. D. Arntz, W. K. den Otter, H. H. Beeftink, R. M. Boom, and W. J. Briels,
{\it{Granular Matter}} {\bf 15}, 25 (2013).




 
 
 


\end{thebibliography}
\end{document}


\alnumsecstyle{Lal}

{\bf Convection of mono-disperse particles in a highly filled rotating cylinder}\\
\\
Shoichi Yoneta, Hiroyuki Ebata, and Shio Inagaki \\
{\it Dept. of Physics, Kyushu University} \\

\section{Derivation of the advection diffusion equations}
The density, flux, and velocity of the total particle system are denoted as $\rho$, $\vb*{j}$, and $\vb*{v}$, respectively. For the colored particle system, we employ the notations $\rho_A$, $\vb*{j_A}$, $\vb*{v_A}$.
The advection equation is derived from the mass conservation law in the colored particle system, which can be written as:
\begin{equation*}
\partial_t\rho_A+\vb*{\nabla}\cdot\vb*{j_A}=0.
\end{equation*}
Here, the flux of the colored particles is given by $\vb*{j_A} = \rho_A\vb*{v_A}-D\vb*{\nabla}\rho_A$. The diffusion coefficient is denoted as $D$. Thus, the mass conservation law can be expressed as:
\begin{eqnarray*}
\pdv{\rho_A}{t}+\vb*{\nabla}\cdot\qty(\rho_A\vb*{v_A}-D\vb*{\nabla}\rho_A)=0.
\end{eqnarray*}
In this study, we assume that the experiment is passive scalar transport of a two-component system of white and colored particles. Passive scalar transport refers to a transport phenomenon, where the concentration of colored particles follows the velocity field without exerting any influence on it.
By considering passive scalar transport, where $\vb*{v} = \vb*{v_A}$, and the incompressibility of the flow $\qty(\vb*{\nabla}\cdot\vb*{v} = 0)$, we can derive the following equation:
\begin{eqnarray*}
\pdv{\rho_A}{t}+\vb*{v}\cdot\qty(\vb*{\nabla}\rho_A)-D\nabla^2\rho_A=0.
\end{eqnarray*}
Furthermore, in the case of the incompressible material $\qty(\partial_t\rho = 0,\hspace{5pt}\vb*{\nabla}\rho = 0)$,
we obtain the advection-diffusion equation for the concentration of the colored particles, $\phi_A$, defined as the ratio of the density of colored particles, $\rho_A$, to the density of the total particle system, $\rho$. The equation is given by:
\begin{eqnarray*}
\pdv{\phi_A}{t}+\vb*{v}\cdot\qty(\vb*{\nabla}\phi_A)-D\nabla^2\phi_A=0.
\end{eqnarray*}
To be precise, the diffusion coefficient is a tensor and can be written as $D\nabla^2\phi_A = D_{ij}\vb*{\nabla_j}\cdot\qty(\vb*{\nabla_i}\rho_A)$.
Applying vector operations in the cylindrical coordinate system, the advection-diffusion equation can be expressed as:
\begin{eqnarray*}
\pdv{\phi_A}{t}+v_r\pdv{\phi_A}{r}+\dfrac{v_\theta}{r}\pdv{\phi_A}{\theta}+v_x\pdv{\phi_A}{x} = \dfrac{1}{r}\pdv{}{r}\qty(D_{rr} r\pdv{\phi_A}{r})+\dfrac{D_{\theta\theta}}{r^2}\pdv[2]{\phi_A}{\theta}+D_{xx}\pdv[2]{\phi_A}{x},
\end{eqnarray*}
where $D_{xx}$, $D_{rr}$, and $D_{\theta \theta }$ are the diagnal components of $D_{ij}$.
By considering axisymmetry $\qty(\partial_\theta = 0)$, the equation can be simplified as:
\begin{eqnarray*}
\pdv{\phi_A}{t}+v_r\pdv{\phi_A}{r}+v_x\pdv{\phi_A}{x} = \dfrac{1}{r}\pdv{}{r}\qty(D_{rr} r\pdv{\phi_A}{r})+D_{xx}\pdv[2]{\phi_A}{x}.
\end{eqnarray*}
Note that $D_{xx}$ is independent of $x$, while $D_{rr}$ depends on $r$, as the diffusion coefficient differs between the surface layer and the inner layer.
In the main article, 
the diffusion coefficients in axial and radial directions are presented by $D_x$ and $D_r$, respectively, instead of $D_{xx}$ and $D_{rr}$.

\section{Fitting function for $v_x^s$}
The axial velocity in the surface layer, $v_x^s$, was fitted to the experimental data using the following function:
\begin{equation*}
 v_x^s = \sum_{k=1}^4 \Big[ A_k \sin 2 k \pi \tilde{x} + B_k \cos (2 k-1) \pi \tilde{x} \Big],
\end{equation*}
where $\tilde{x}=x/L - 1/2$.
This involves adding terms with sine functions having phases that are even multiples of $\pm \pi/2$ and terms with cosine functions having phases that are odd multiples of $\pm \pi/2$ at $\tilde{x}=\pm 1/2$. This combination ensures that the values at both ends of the cylinder are zero.
Here is the list of the fitting parameters of $A_k$ and $B_k$.

\begin{table}[h]
	\begin{center}
		\begin{tabular}{|c|r|r|r|r|} \hline
		$k$ & 1 & 2 & 3 & 4 \\ \hline\hline
		$A_k$ & 42.1 & -15.0 & 4.2 & -1.7\\ \hline
		$B_k$ & 8.1 & -5.3 & 2.5 & -1.0\\ \hline
	\end{tabular}
	\caption{Fitting parameters. Unit is mm/h.}
	\end{center}
\end{table}

\section{Derivation of the radial velocity, $v_r$}
Applying vector operations in the cylindrical coordinate system, the incompressible flow can be described as follows:
\begin{eqnarray*}
\vb*{\nabla}\cdot\vb*{v} = \dfrac{1}{r}\pdv{}{r}\qty(rv_r)+\dfrac{1}{r}\pdv{v_\theta}{\theta}+\pdv{v_x}{x} = 0.
\end{eqnarray*}
Under the assumption of axial symmetry ($\partial_\theta = 0$) and the $r$-independence of $v_x$ within each layer, the radial velocity takes on a functional form:
\begin{equation} \label{eq:incompressibleFlow}
	\begin{dcases}
		v_r^i(x,r) = -\dfrac{r}{2}\dv{v_x^i}{x}+\dfrac{C^i(x)}{r}, \hspace{15pt}\qty(r_i\leq r\leq r_b)  \\
		v_r^s(x,r) =-\dfrac{r}{2}\dv{v_x^s}{x}+\dfrac{C^s(x)}{r},\hspace{15pt}\qty(r_b\leq r\leq r_s)
	\end{dcases}
\end{equation}
where $C^\alpha(x)$ is the integration constants, and the superscript letter $\alpha$ represents $i$ for the inner layer and $s$ for the surface layer. 
By considering the mass jump condition \cite{Chadwick99:massJumpCondition}, we ensure the continuity of the normal component of advection at the boundary between layers at $r=r_b$. 
Since advection is the product of density and velocity, and we assume the material to be incompressible, there is no density change between the surface and inner layers. Consequently, the radial velocity, which represents the normal component of velocity, remains continuous at the layer boundary. This can be expressed mathematically as:
\begin{equation} \label{eq:massJump}
	-\dfrac{r_b}{2}\dv{v_x^i}{x}+\dfrac{C^i(x)}{r_b} = -\dfrac{r_b}{2}\dv{v_x^s}{x}+\dfrac{C^s(x)}{r_b}.
\end{equation}
From this equation, we can deduce:
\begin{equation} \label{eq:massJumpComdition}
C^i(x)-C^s(x) = \dfrac{{r_b}^2}{2}\qty(\dfrac{dv_x^i}{dx}-\dfrac{dv_x^s}{dx}).
\end{equation}
\begin{figure}[t]
	\begin{center}
	 \includegraphics[width=6cm]{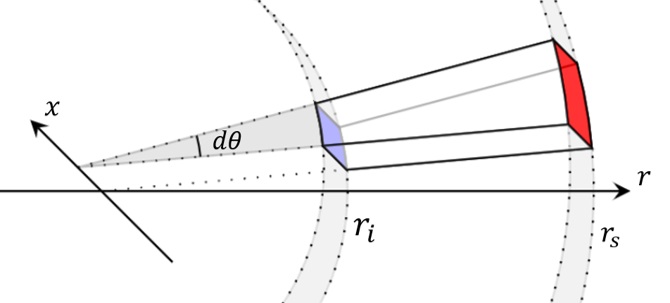}
	 \caption{
	 Volume element of a coaxial cylinder in the cylindrical coordinates.} \label{fig:dV}
	\end{center}
\end{figure}
These integration constants can be determined by the principle of flow conservation.
Now, we consider a volume element named $\delta V$ within the cylindrical coordinate system, as depicted in Fig. \ref{fig:dV}.
This volume is bounded by radial distances $[r_i, r_s]$, axial positions  $[x', x'+\delta x]$, and azimuthal angles $[\theta', \theta'+\delta \theta]$. Here, $-L/2 \le x' \le  L/2$ and $0 \le \theta'< 2 \pi$ are arbitrary values, and  $\delta x$ and $\delta \theta$ represent infinitesmal increments in the $x$ and $\theta$ coordinates, respectively. 
Assuming media incompressibility entails a constant density and the absence of inflow and outflow. Consequently, when these conditions are met, the net flux through any given cross-section is required to be zero. Therefore, among the six faces of the volume element, the two faces located at $x=x'$ and $x=x'+\delta x$, which are perpendicular to the axis, must also have a flux of zero.
Furthermore, the assumption of axial symmetry implies that there is no flow in the azimuthal direction.
Therefore, the inflow and outflow of particles in the axial direction must be balanced with that in the radial direction.
The net flux through the surface of this volume element can be expressed as follows:
\begin{eqnarray*}
\displaystyle\int_{\theta'}^{\theta'+d\theta}\int_{x'}^{x'+dx}v_x^i(r_i,x)r_i d\theta dx + \displaystyle\int_{\theta'}^{\theta'+d\theta}\int_{x'}^{x'+dx}v_x^s(r_s,x)r_sd\theta dx = 0.
\end{eqnarray*}
Since the integral ranges of the two terms are identical, the following equation holds:
\begin{equation} \label{eq:vrPreservation}
C^i(x)+C^s(x) = \dfrac{1}{2}\qty({r_i}^2\dfrac{dv_x^i}{dx}+{r_s}^2\dfrac{dv_x^s}{dx})
\end{equation}
By combining equations (\ref{eq:incompressibleFlow}), (\ref{eq:massJumpComdition}), and (\ref{eq:vrPreservation}), we can determine the radial velocity, $v_r$, as follows:
\begin{eqnarray*}
    \begin{dcases}
			v_r^i(x,r) =\dfrac{{r_i}^2-r^2}{2r}\dv{v_x^i(x)}{x},    &   (r_i\leq r\leq r_b)  \\
      v_r^s(x,r) =\dfrac{{r_s}^2-r^2}{2r}\dv{v_x^s(x)}{x}.    &   (r_b\leq r\leq r_s)
    \end{dcases}
\end{eqnarray*}

\section{Numerical simulation of the advection-diffusion model}
To validate our advection-diffusion equations, we numerically solved them by using an explicit finite difference method.
The discretization of the equation involved applying the cubic-interpolated pseudo-particle (CIP) method for the advection term \cite{Yabe87:CIP}, while the forward Euler method was used for the remaining terms. In our implementation, we employed a staggered grid for the computational points and applied the Neumann boundary condition. The radial, axial, and temporal steps are represented by the discretized parameters $\Delta r$, $\Delta x$, and $\Delta t$, respectively. We have chosen $\Delta r = \Delta x = 1.0 \, \rm{mm}$ and $\Delta t = 0.0025 \, \rm{h}$. 

\section{Effect of rotational speed of a cylinder}
\begin{figure}[h]
	\begin{center}
	 \includegraphics[width=8cm]{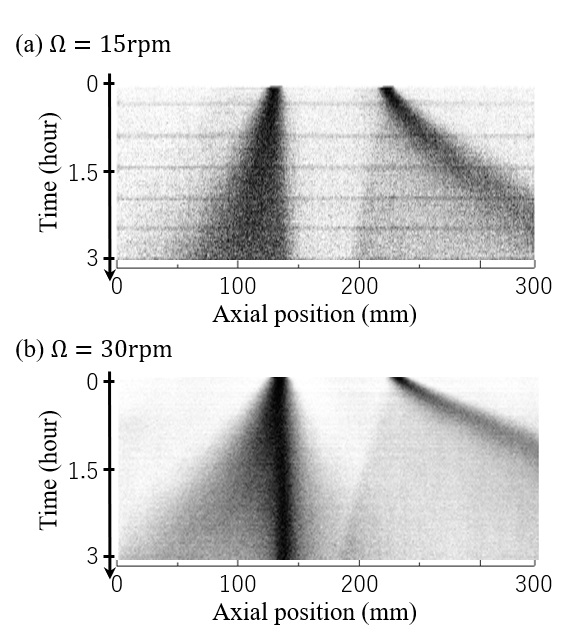}
	 \caption{
		Spatio-temporal diagrams with two pulses.
		(a) $\Omega=15 \, \rm{rpm}$
		(b) $\Omega=30 \, \rm{rpm}$ 
		The fill level is fixed at 95 $\%$.
		} \label{fig:Diff}
	\end{center}
\end{figure}

Based on the advection diffusion equations, we can estimate the diffusion coefficient, $D_x^s$, from the experimental data.
Figure \ref{fig:Diff} shows the spatio-temporal diagram of two pulses. By measuring how the pulses blurred as they moved axially, we can estimate the axial diffusion coefficient in the surface layer.
However, in addition to the diffusive effect of the surface avalanche, the radial emanation of colored particles contributes to the blurring process of the pulses. Consequently, when estimating the axial diffusion coefficient from the spatio-temporal diagram, it is anticipated that the estimated value will appear larger than the actual value.
In fact, the diffusion coefficient used in the numerical calculations is $30 \, \rm{mm^2/s}$, which is roughly $82 \, \rm{mm^2/s}$ when estimated with experimental data.
Therefore, the diffusion coefficient in the surface layer, $D_x^s$, was chosen carefully so that the numerical solution matches the experimental results.
\\
The spatio-temporal diagram is not completely symmetric due to the filling process. When we confined the particles, we placed the cylinder vertically. After pouring all the particles, we sealed the end, then placed the cylinder horizontally. During the filling process, we manually tapped to achieve dense configuration so that we can minimize the gradient of the packing density along the axis as much as possible. The